\documentstyle[prl,aps,preprint]{revtex}
\newcommand{\br}{{\bf r}}
\newcommand{\bu}{{\bf u}}
\newcommand{\bq}{{\bf q}}

\newcommand\ignore[1]{}

\begin{document}

\draft
\title{
Stability of the vortex lattice in $D$-wave superconductors}
\author{Jun'ichi Shiraishi${}^1$,
Mahito Kohmoto${}^1$ and Kazumi Maki${}^2$}
\address{${}^1$Institute for Solid State Physics,
University of Tokyo, Roppongi, Minato-ku, Tokyo 106, Japan}
\address{${}^2$Department of Physics and Astronomy,
University of Southern Calfornia Los Angeles,
Cal. 90089-0484, USA}

\maketitle
\begin{abstract}
Use is made of Onsager's hydrodynamic equation to derive the vibration
spectrum of the vortex lattice in $d$-wave superconductor.
In particular the rhombic lattice ({\it i.e.}
the $45^\circ$ tilted square lattice)
is found to be stable for $B>H_{cr}(t)$.
Here $H_{cr}(t)$ denotes the critical field at which the
vortex lattice transition takes place.
\end{abstract}
\pacs{ }

\narrowtext

Since the discovery of the triangular vortex lattice
in type II superconductors by Abrikosov\cite{1}
and others\cite{2,3},
the vibrational modes of the vortex lattice have been studied by a
number of people\cite{4}.
In these works the stability of the triangular vortex lattice in an
$s$-wave
superconductor is established.

The discovery of hole-doped
high $T_c$ cuprate superconductivity by
Bednorz and M\"uller\cite{5}
in 1986 and the recent realization\cite{6}
that $d$-wave superconductivity is involved may
give a new twist on the whole subject.

We have shown earlier that the
rhombic vortex lattice
(or the $45^\circ$ tilted square lattice)
is stable in the vicinity of the upper critical field in a $d$-wave
superconductor in a magnetic field parallel to the $c$ axis\cite{7}.
More recently we have shown \cite{8}
the vortex lattice transition from the triangular lattice to
the square lattice takes
place at a small magnetic field $H_{cr}(t)\sim \kappa^{-1} H_{c2}(t)$,
which implies that the vortex lattice should be rhombic in the
overwhelming region in the $B$-$T$ phase diagram
(see Fig. 1).
Here $\kappa$ is the Ginzburg-Landau parameter and we
have $H_{cr}$ about a few Tesla in YBCO and Bi2212
at low temperatures.
Indeed this critical field $H_{cr}(t)$
is consistent with the observation of
rhombic vortex lattice by SANS\cite{9}
and by STM imaging\cite{10}
in monocrystals of YBCO at 3 Tesla.

The object of this paper is to study the
vibrational spectrum of the $45^\circ$
tilted
square vortex lattice in a magnetic field. First following Fetter
et al.\cite{4}, we study Onsager's
and Landau's hydrodynamic equation\cite{11,12}
for a vortex lattice.
Basically we assume that the vortex moves with the local velocity
generated by other vortices.
We find that the square vortex lattice
is unstable when $B<H_{cr}(t)$ but
becomes stable
for  $B\geq H_{cr}(t)$.
The vibrational spectrum are determined in the
whole Brillouin zone.
Second we analyze the vibration spectrum within the time dependent
Ginzburg-Landau equation\cite{13} with the
Aranov-Hikami-Larkin term\cite{14}.
In this limit we have now a set of damped oscillation modes rather
than oscillation modes.
However the stability condition of the square vortex lattice is
the same as in the analysis using Landau's
hydrodynamic equation as expected.
For example this new vibration spectrum will be crucial in determining
the melting transition line
where the vortex lattice melts into a vortex liquid.
\bigskip

\noindent
{\bf I. Vibration modes in the
square lattice (hydrodynamic limit)}\\
\noindent
As is well known the Landau's and Onsager's hydrodynamic equation
applies when motion of vortices involves no energy dissipation.
Unfortunately this condition is never realized
for vortices in a $d$-wave superconductor since
there are the low energy extended states
attached to every vortex and they certainly
dissipate energy whenever the vortex is
in motion \cite{15}.
Nevertheless it is of great interest to
study this idealized limit.

The extended Ginzburg-Landau free energy for
the vortex state with $\xi\ll d \ll \lambda$ in $d$-wave
superconductors reduces to \cite{8}
\begin{eqnarray}
\Omega&=&{2\pi n_\phi \xi^2 \over \kappa^2}
\sum_L{}' \phi(\br_L-\br_0), \label{GL}\\
\phi(\br)&=&
-\epsilon a_2 \xi^2 \kappa^2
{\cos 4 \theta_{\br} \over |\br|^4}+
K_0\left(r\over \lambda  \right) ,
\end{eqnarray}
where
$
\epsilon= {31\zeta(5)(-\ln t)/ 196 \zeta(3)^2} \sim 0.114(-\ln t),
$
$t=T_c/T$,
$a_2={20\over 3} (\ln 2 +{ 1\over 8})=5.454\cdots$\footnote{
Note that $a_2$ is related to the constant $a_1$ given
in Ref. \cite{8}
as $a_2=5 a_1/2\pi$.}; and $\xi,d$,
and $\lambda$ are the coherence length, intervortex
distance and the magnetic penetration depth respectively.
Here $\phi(\br)$ describes the interaction energy between two
vortices; the first term is the core interaction
(new to $d$-wave superconductors)
while the second term is the usual magnetic interaction.
Following Onsager\cite{12} the equation for the $L$-th
vortex is
\begin{eqnarray}
\dot{\bf u}_L= -
{\bar{\kappa} \over 2\pi}
(\hat{{\bf z}}\times \nabla)
\sum_{L'\neq L}
({\bf u}_{LL'}\cdot \nabla)
\;\phi(\br_{L'}-\br_L),
\label{eq1}
\end{eqnarray}
where $\bar{\kappa}={h / 2m}={\pi / m}$ (in quantum unit),
${\bf r}_L$ is the position of the $L$-the vortex,
${\bf u}_L$ is the small displacement from
the equilibrium position ${\bf r}_L^0$
(namely ${\bf r}_L={\bf u}_L+{\bf r}_L^0$),
$\nabla$ denotes the differentiation with respect to
${\bf r}_L$.
We have set
${\bf u}_{LL'}={\bf u}_L-{\bf u}_{L'}$,
${\bf r}_{LL'}={\bf r}_L-{\bf r}_{L'}$.
In deriving Eq.(\ref{eq1}) the Hamiltonian $H({\bf r}_L)$
in Onsager's theory is replaced
by the free energy $\Omega({\bf r}_L)$
obtained in Ref. \cite{8}.
In Ref. \cite{8}, the square vortex lattice tilted
$45^\circ$ was studied.
In the following analysis, however, it is more convenient to use the
coordinates system rotated by $\pi/4$ around the $z$ axis,
so that vortex lattice axes coincide with the
$x$- and $y$-axes. This change makes the sign in front of the
core interaction term opposite from the one in Ref. \cite{8}.
The equilibrium positions of the square vortex lattice
become ${\bf r}_L^0=d (m,n)$
($m,n\in {\bf Z}$),
where $d=\sqrt{\phi_0/B}$ ($\phi_0$ is the flux quantum).

Now introducing a plane wave representation as
${\bf u}_L= {\bf s} \exp (i({\bf q}\cdot {\bf r}_L^0 -\omega t))$,
Eq.(\ref{eq1}) is rewritten as
\begin{eqnarray}
&&-i\omega s_x = \alpha s_x + \beta s_y,\\
&&-i\omega s_y = \gamma s_x - \alpha s_y,\nonumber
\end{eqnarray}
where
\begin{eqnarray}
\alpha &=&
{\bar{\kappa} \over 2\pi\lambda^2 }
\sum_L{}'
(1-e^{i {\bf q}\cdot {\bf r}_L})
\left[
{x_L y_L \over r_L^2 }
K_2\left(r_L\over \lambda \right)
-20 \epsilon a_2 \kappa^2 \xi^4 \lambda^2
 {\sin 6\theta_L \over r_L^6}
\right],\\
\xi &=&{\beta+\gamma \over 2}=
{\bar{\kappa} \over 4\pi\lambda^2 }
\sum_L{}'
(1-e^{i {\bf q}\cdot {\bf r}_L})
\left[
{y_L^2- x_L^2 \over r_L^2}K_2\left(r_L\over \lambda \right)
+40 \epsilon a_2 \kappa^2 \xi^4  \lambda^2 {\cos 6\theta_L \over r_L^6}
\right],\\
\eta &=&{\beta-\gamma \over 2}=
{\bar{\kappa} \over 4\pi\lambda^2}
\sum_L{}'
(1-e^{i {\bf q}\cdot {\bf r}_L})K_0\left(r_L\over \lambda \right),
\end{eqnarray}
where $K_0(x)$ and $K_2(x)$ are the modified Bessel functions.
The vibration frequency $\omega$ is then given by
\begin{eqnarray}
\omega^2=-\alpha^2-\beta\gamma=\eta^2-\alpha^2-\xi^2.
\end{eqnarray}
Let us consider the long wave length limit
($q=|{\bf q}| \ll 1$) of the vibration
spectrum.
Then we obtain
\begin{eqnarray}
-\xi+i \alpha
&=&
{\bar{\kappa} \over 32 \pi \lambda^2} q^2 d^2
\left[
(\Sigma_{24}- {a(B)\over \mu^{2}}\Lambda_8) e^{2 i\chi}+
(\Sigma_{24}-8 \Sigma_{xy4}-
{a(B)\over \mu^{2}}\Lambda_4) e^{-2 i\chi}\right],\\
\eta&=&
{\bar{\kappa} \over 16 \pi \lambda^2} q^2 d^2 \Sigma_1,
\end{eqnarray}
where $\chi$ is the angle ${\bf q}$ makes with the $x$ axis, we have set
\begin{eqnarray}
a(B)={40 \epsilon a_2 \kappa^2 \xi^4 \over \phi_0^2}B^2
\simeq 0.145025 \left(B\over H_{cr}\right)^2,
\end{eqnarray}
where $\mu=d/\lambda( \ll 1)$ and
\begin{eqnarray}
H_{cr}(t)\equiv \phi_0\sqrt{4\pi C-1 \over 10 \epsilon a_2 \kappa^2 \xi^4
(\Lambda_8+\Lambda_4)}
\simeq 0.48 (-\ln t)^{-1/2} {H_{c2}(t) \over \kappa}.
\label{hcr}
\end{eqnarray}
We have also used the notations
\begin{eqnarray}
&&\Sigma_1=d^{-2} \sum_L{}' r_L^2 K_0(r_L/\lambda)
={8\pi \over \mu^4}+ O(1),\\
&&\Sigma_{24}=d^{-2} \sum_L{}' r_L^2 K_2(r_L/\lambda)
={16 \pi \over \mu^4}-{2 \over \mu^2}+O(1),\\
&&\Sigma_{xy4}=d^{-2} \sum_L{}' {x_L^2 y_L^2 \over r_L^2}
 K_2(r_L/\lambda)
={2\pi \over \mu^4} -C {2\pi  \over \mu^2} + O(1),\\
&&\Lambda_{8}=d^{4}\sum_L{}' {1 \over r_L^4} \cos(8 \theta_L)
=5.0306\cdots,\\
&&\Lambda_{4}=d^{4}\sum_L{}' {1 \over r_L^4} \cos(4 \theta_L)
=3.1512\cdots,
\end{eqnarray}
where $C=0.10331$.
The quantities $\Sigma_1,\Sigma_{24}$, $\Sigma_{xy4}$ and $C$ are
quoted from \cite{4} for the sake of the readers convenience.
Using these, we have
\begin{eqnarray}
\omega^2\left({32 \pi\ \over \bar{\kappa} q^2 \mu^2 }\right)^2
&=&
={32 \pi \over \mu^6}
\Bigl[
(2+a(B)\Lambda_8)-(16\pi C-2 - a(B) \Lambda_4) \cos 4\chi
\Bigr]+O(\mu^{-4}).
\end{eqnarray}
Finally, we obtain
\begin{eqnarray}
\omega^2
&\simeq&
\left( eB \over mc\right)^2 q^4 \lambda^2 d^2
\Biggl[\left(0.01989+0.007257 \left( B\over H_{cr} \right)^2 \right)
 \nonumber \\
&&\qquad\qquad \qquad\qquad
-\left(0.03176-0.004546 \left( B\over H_{cr} \right)^2 \right)
\cos 4\chi\Biggr].
\end{eqnarray}
The condition to
have real spectrum ($\omega^2>0$) is $B\geq H_{cr}$.
We see immediately that the square vortex lattice is unstable for
$B<H_{cr}$ for some $\chi$,
while it becomes stable for all $\chi$ for $B>H_{cr}$
as it should be.
Exactly at the transition point $B=H_{cr}$
$\omega^2=0$ at $\chi=0,\pi/2,\pi$ and $3\pi/2$.

In the previous paper \cite{8}, we derived the critical
field as $H_{cr}(t)\simeq 0.52 (-\ln t)^{-1/2} H_{c2}(t)/ \kappa $
by minimizing the free energy of the extended Landau Ginzburg theory.
This and the result obtained here in Eq. (\ref{hcr})
appear to be fully consistent.
\bigskip

\noindent
{\bf II. Thermal fluctuation of the vortex lattice}\\
\noindent
Let us consider thermal fluctuation of the square
vortex lattice ($B>H_{cr}$) based on the
Landau-Ginzburg free energy (\ref{GL}).
This provides us with another
way to study the
quantities $\alpha,\beta$ and $\gamma$ introduced
in the last section.

The small deviation of the free energy $\Delta \Omega$ due to the
lattice vibration is
\begin{eqnarray}
\Delta \Omega= {\pi n_\phi \xi^2 \over \kappa^2}
\sum_{L}{}'
\sum_{\mu,\nu=x,y}
(u_{L\mu}-u_{0\mu}) \phi_{\mu\nu}(\br^0_L) (u_{L\nu}-u_{0\nu}),
\end{eqnarray}
where $u_{L\mu}$ denotes the $\mu$-component of the
small deviation vector $\bu_L$ and
$\phi_{\mu\nu}(\br)= \partial_\mu\partial_\nu \phi(\br)$.
Introducing the Fourier transformation
$u_{L\mu}=\sum_{\bq}e^{i \bq \cdot \br^0} s_\mu(\bq)$
(the reality condition $s_\mu(-\bq)=s_\mu(\bq)^*$ is imposed),
we have
\begin{eqnarray}
\Delta \Omega&=&{2\pi n_\phi \xi^2 \over \kappa^2}
\sum_{L}{}' \sum_{\mu,\nu,\bq}
s_\mu(\bq)^* s_\nu(\bq) (1-e^{i \bq \cdot
  \br^0})\phi_{\mu\nu}(\br_L^0)\nonumber\\
&=&
{4 \pi^2 n_\phi \xi^2 \over \kappa^2 \bar{\kappa}}
 \sum_{\bq}
\Bigl(s_x(\bq)^*,s_y(\bq)^*\Bigr)
\left(\begin{array}{cc}
-\gamma & \alpha \\ \alpha&\beta \end{array} \right)
\left(\begin{array}{c}
s_x(\bq) \\ s_y(\bq) \end{array} \right) \nonumber\\
&=&
{4 \pi^2 n_\phi \xi^2 \over \kappa^2 \bar{\kappa}}
 \sum_{\bq}
\left(  \omega_+|s_+(\bq)|^2 +\omega_-|s_-(\bq)|^2  \right),
\end{eqnarray}
where $s_\pm(\bq)$ denotes a suitable unitary transformation of
$s_\mu(\bq)$ and
\begin{eqnarray}
&&\omega_\pm=\eta\pm \sqrt{\alpha^2+\xi^2} \nonumber \\
&=&
\left( {\bar{\kappa} \over 32 \pi \lambda^2} q d^2 \right)
\left[
{16\pi \over \mu^4}\pm
\left(
{16\pi \over \mu^4} -
{(2+a(B) \Lambda_8)-(16\pi C -2- a(B)\Lambda_4) \cos 4 \chi \over \mu^2}
\right)
\right] \label{omegapm}\\
&=&
\left(eB\over mc\right) q^2 \lambda^2
\Biggl[
{1\pm 1\over 2} \mp
\left(d\over \lambda\right)^2
\Biggl( \left(
   0.01989+0.007257 \left(B\over H_{cr}\right)^2
       \right) \nonumber \\
&& \qquad\qquad\qquad
-
 \left(
   0.03176+0.004546 \left(B\over H_{cr}\right)^2
       \right) \cos 4\chi
\Biggr)
\Biggr].
\end{eqnarray}
The stability condition $\omega_\pm>0$ leads to
$\eta> \sqrt{\alpha^2 + \xi^2}$ or $B>H_{cr}$.
Note that this condition is exactly the same as
the one in the previous section.
Evaluating the path integral
$\int {\cal D}s\; (|s_x|^2+|s_y|^2) e^{-\beta \Omega} /
\int {\cal D}s\; e^{-\beta \Omega}$, we obtain
\begin{eqnarray}
\sum_\bq \langle |s_x(\bq)|^2+|s_y(\bq)|^2\rangle=
kT { \kappa^2 \bar{\kappa}\over 4 \pi^2 n_\phi \xi^2 }
\sum_\bq \left( {1\over \omega_+(\bq)}+ {1\over \omega_-(\bq)}\right).
\end{eqnarray}
Studying this expression we will be able to study the
vortex lattice melting from the square vortex lattice.
Details on the vortex lattice melting
will be reported in a separate paper\cite{12.5}.
\bigskip

\noindent
{\bf III. Time dependent Ginzburg Landau equation approach}\\
\noindent
In a $d$-wave superconductor the vortex motion cannot be purely
hydrodynamic,
since the vortex motion always accompanied
with the energy dissipation as seen from a number of experiments
involving the flux
flow resistance in high $T_c$ cuprate superconductors.
Therefore the vibration models for the vortex lattice should be
described better by the time-dependent Ginzburg Landau
equation\cite{13}, supplemented by the term which accounts for the Hall
effect
\cite{14} at least qualitatively.

In the vicinity of $T=T_c$ the time dependent Ginzburg Landau (TDGL)
equation is the same as that in $s$-wave superconductors\cite{13}.
Neglecting the term which gives rise to the Hall effect we obtain
\begin{eqnarray}
{\pi \over 8 T_c}{\partial \over \partial t}
\Delta({\bf r})=N_0^{-1}
{\partial \Omega \over \partial \Delta^*({\bf r})},\label{tdgl}
\end{eqnarray}
where $\Omega$ is the free energy defined in Ref. \cite{8}.
Assuming the usual product expression
$\Delta({\bf r})=\Delta \prod_i f({\bf r}-{\bf r}_i)$, we obtain
\begin{eqnarray}
{\pi \over 8 T_c}\Delta^2 {\bf v}_i {1 \over 2}
\left|{\partial f({\bf r}-{\bf r}_i)\over \partial {\bf r}_i}
\right|^2
\prod_{j\neq i}
|f({\bf r}-{\bf r}_i)|^2
=N_0^{-1}
{\partial \Omega \over \partial {\bf r}_i},
\end{eqnarray}
which reduces to
\begin{eqnarray}
{\bf v}_i={16 T_c \over \pi} {\xi^2(T) \over \Delta^2(T)}
N_0^{-1} {\partial \Omega \over \partial {\bf r}_i}. \label{vi}
\end{eqnarray}
The right hand side of Eq.(\ref{vi}) is evaluated as in
the case of the hydrodynamic limit, and obtain two frequencies
\begin{eqnarray}
 \omega_\pm=-i A_0 (\eta\pm \sqrt{\xi^2+\alpha^2}),\label{ompm}
\end{eqnarray}
where $\eta\pm \sqrt{\xi^2+\alpha^2}$
is calculated in Eq. (\ref{omegapm}) and
\begin{eqnarray}
A_0={14\zeta(3) \over \pi^3} {E_F\over T_c}.
\end{eqnarray}
The stability of the vortex lattice requires
$\eta >\sqrt{\xi^2+\alpha^2}$, which is satisfied for
$B>H_{cr}(t)$.

Finally the gauge invariance of TDGL requires that Eq.(\ref{tdgl})
has to be replaced by
\begin{eqnarray}
{\pi \over 8 T_c}(1-i\sigma){\partial \over \partial t}
\Delta({\bf r})=N_0^{-1}
{\partial \Omega \over \partial \Delta^*({\bf r})},\label{tdgl2}
\end{eqnarray}
where
\begin{eqnarray}
\sigma={4T_c \over \pi} {\partial \ln T_c \over \partial \mu}.
\end{eqnarray}
Then in the presence of the Hall effect
Eq.(\ref{ompm}) should be replaced by
\begin{eqnarray}
\omega_\pm={\sigma-i\over 1+\sigma^2}
 A_0 (\eta\pm \sqrt{\xi^2+\alpha^2}).\label{ompm2}
\end{eqnarray}
Therefore in this general situation the vibration becomes damped
oscillation. However, the
stability condition is the same as before.

\noindent
{\bf Concluding Remarks}\\
\noindent
Limiting ourselves to the square vortex lattice,
we have studied the oscillation mode of the vortex lattice first
by the hydrodynamic
equation and then by the time-dependent
Ginzburg Landau equation.
Both analysis indicates that
the square vortex lattice is stable for $B>H_{cr}(t)$,
while it becomes unstable for $B<H_{cr}(t)$.
Of particular interest is that the fluctuation of the
vortex lattice diverges linearly like $(B-H_{cr}(t))^{-1/2}$
as $B$ approaches the transition point. This should have a
profound implication on the melting of the vortex lattice for example.
The related problems will be discussed in future publications.

\begin{center}
{\large Acknowledgment}
\end{center}

One of us (KM) would like to thank CREST enabled him to visit two
weeks at ISSP, University of Tokyo. This work is supported by NSF
under grant number DMR 95-31720.

\begin{figure}

Fig. 1
The $B$-$T$ phsae diagram.

\end{figure}

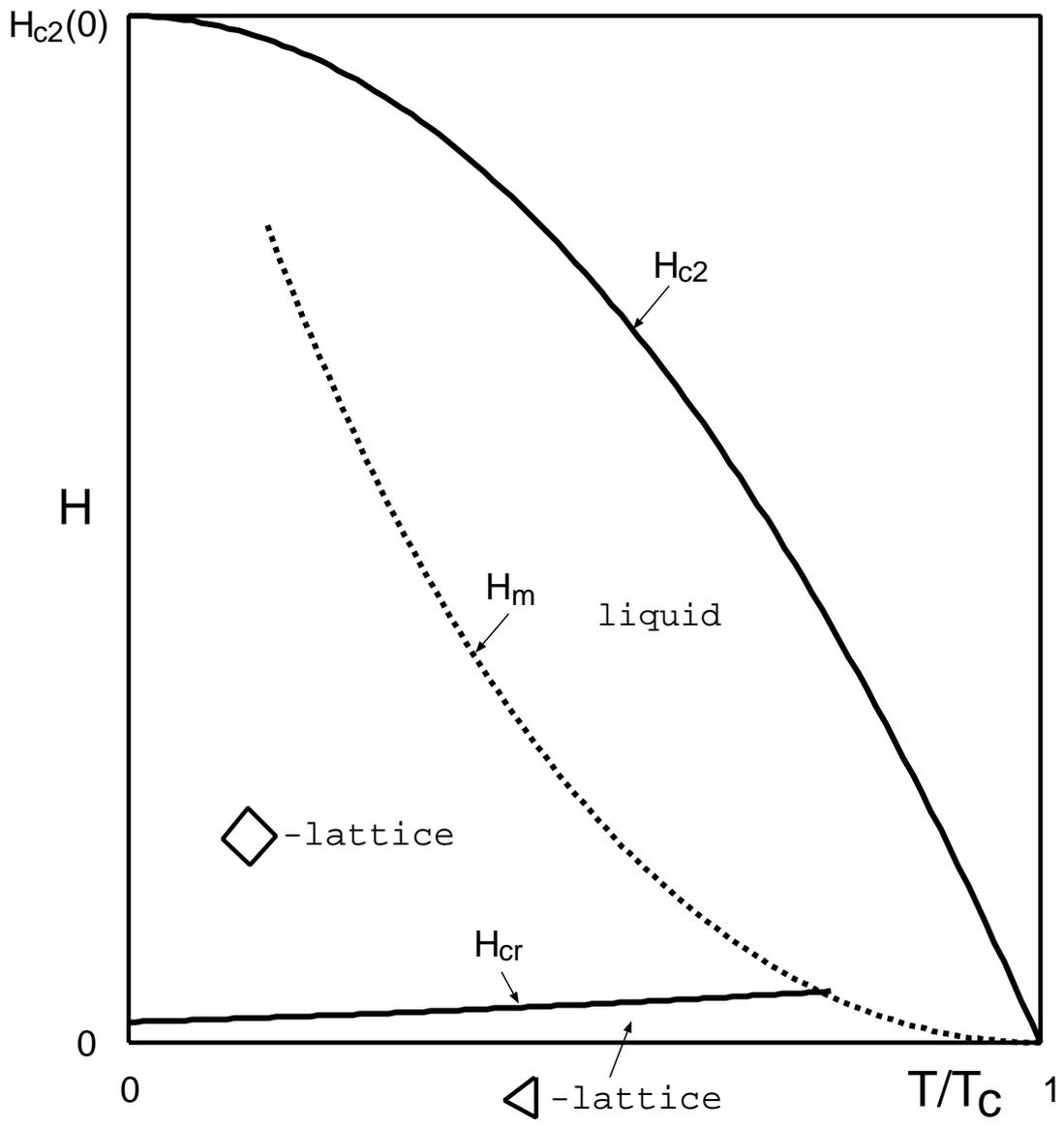
\end{document}